\documentclass[conference]{IEEEtran}

\usepackage{float}
\usepackage{enumitem}
\usepackage{cite}
\usepackage{graphicx,amsmath,amssymb,url,multirow,subcaption}

\makeatletter

\begin{document}

\title{Closed-Form Solutions for a Low-Order System Fast Frequency Response Model}

\author{\IEEEauthorblockN{Julius Susanto\IEEEauthorrefmark{1}, Alireza Fereidouni\IEEEauthorrefmark{1}, Pierluigi Mancarella\IEEEauthorrefmark{2}, and Dean Sharafi\IEEEauthorrefmark{1}}
\IEEEauthorblockA{\IEEEauthorrefmark{1}Department of Power System \& Market Planning, Australian Energy Market Operator, Perth, WA 6000, Australia}
\IEEEauthorblockA{\IEEEauthorrefmark{2}Department of Electrical and Electronic Engineering, The University of Melbourne, Melbourne, VIC 3010, Australia}
}

\maketitle

\begin{abstract}
This paper presents a novel closed-form solution for a low-order system frequency response (SFR) model that is accurate for all time periods and an accompanying approximation for representing primary frequency responses at two different speeds while still maintaining mathematical tractability. This allows for the inclusion of both fast frequency responses (e.g. from battery energy storage systems) and more conventional responses (e.g. from thermal generation) in a single SFR formulation. The closed-form expressions can be efficiently used in applications that use the SFR model such as frequency stability studies and security-constrained unit commitment (SCUC) studies.
\end{abstract}

\vspace{1em}


\section{Introduction}

Low-order system frequency response (SFR) models are single-machine equivalent models of a power system that can be used to predict the system frequency trajectory in response to a disturbance \cite{anderson_1990}. The differential equations that underpin the model are typically solved via numerical integration, but sometimes it is desirable to have closed-form solutions, particularly in optimisation applications, e.g. for security-constrained unit commitment and market models \cite{Ahmadi_2014}.

A closed-form solution for a generalised multi-machine SFR model was proposed in \cite{Aik_2006}, based on the assumption that the dominant generation sources in the system were reheat-type steam boilers. The solution used lead-lag representations of the primary frequency response (PFR) providers and was developed by taking a partial fractions expansion of the combined transfer function, but the frequency nadir cannot be solved analytically. The formulation was simplified in \cite{Ahmadi_2014} by assuming that all turbine reheat time constants were identical. This made the closed-form solution more tractable and allowed for the frequency nadir to be calculated directly. However, this simplification only allows for a single speed response for all PFR providers. In \cite{teng_2015}, a closed-form solution using a ramp response was proposed that can be extended in a tractable way for multiple speeds of response, but there are issues around the accuracy of the solution (discussed in Section II.A below).

The contributions of this paper are threefold: i) introduction of a closed-form solution to a generic low-order SFR model based on a PFR modelled as a lag response, ii) development of an approximation for a lag response PFR with two different speeds of response (or ramp rates), and iii) presentation of novel example applications that use the closed-form expressions derived herein. The inclusion of a second speed of response is crucial in order to incorporate the influence of fast frequency response (FFR) providers, e.g. battery energy storage systems. This is desirable for the design of new frequency control ancillary services markets such as in the Wholesale Electricity Market (WEM) in Western Australia, which is currently undergoing a reform process and is explicitly including the speed of response in the new frequency control markets \cite{etiu_2019}.

The structure of the rest of this paper is as follows: Section II derives the closed-form solutions to the SFR model. An approximation for the SFR model with two speeds of response is proposed in Section III, along with an assessment of the accuracy of the approximation. Example applications for the closed-form solutions are provided in Section IV and finally, Section V offers some conclusions.

\section{Closed-Form Solution for the SFR Model}

A closed form mathematical solution for the SFR model can be derived given certain assumptions: i) PFR is modelled as a continuous, differentiable and monotonically increasing function of time, and ii) demand responses are not modelled as discontinuous active power reductions triggered on the system frequency.

With these assumptions, the linear ordinary differential equation (ODE) for the SFR can be formulated as follows:

\begin{equation}
\label{eqn:SMM_ode}
\frac{d\Delta f(t)}{dt} = \frac{f_n}{2 KE} \left[ p(t) - P_{cont} - D P_{load} \Delta f(t) \right] 
\end{equation}

\noindent where $\Delta f(t)$ is the change in frequency at time $t$ (Hz), $f_n$ is the nominal frequency ($=50$ Hz), $KE$ is the system post-contingency kinetic energy (MW.s), $p(t)$ is the primary frequency response at time $t$ (MW), $P_{cont}$ is the generation contingency size (MW), $D$ is the load relief factor (\% MW/Hz) and $P_{load}$ is the system load at the onset of the contingency (MW).

Denoting $D' = D P_{load}$ and $H = \frac{KE}{f_n}$, (\ref{eqn:SMM_ode}) can be rewritten as follows:

\begin{equation}
\label{eqn:ode_form}
\frac{d\Delta f(t)}{dt} + \frac{D'}{2H} \Delta f(t) = \frac{1}{2H} \left[ p(t) - P_{cont} \right] 
\end{equation}

The general solution to this differential equation is:

\begin{equation}
\label{eqn:ode_form3}
\Delta f(t) e^{\frac{D'}{2H}t} = \frac{1}{2H} \int \left[ p(t) - P_{cont} \right] e^{\frac{D'}{2H}t} dt
\end{equation}

Note that while the formulation above is constructed for under-frequency events, it is also valid for over-frequency events with the contingency $P_{cont}$ as a negative number and the PFR function $p(t)$ as a monotonically decreasing function of time.

\subsection{Linear Ramp Response}

Consider a primary frequency response function of the form $p(t) = \frac{PFR}{t_r} t $ where $PFR$ is is the maximum quantity of PFR delivered (MW) and $t_r$ is the ramp time (s). Denoting $R = \frac{PFR}{t_r}$, then the closed-form solution to (\ref{eqn:ode_form3}) is as follows \cite{teng_2015}:

\begin{equation}
\Delta f(t) = \frac{R t}{D'} - \left( \frac{2 R H}{D'^{2}} + \frac{P_{cont}}{D'} \right) \left( 1 - e^{-\frac{D'}{2H}t} \right)
\end{equation}

\begin{figure}
	\centering
	\begin{subfigure}[t]{0.15\textwidth}
	\centering
        \includegraphics[width=\linewidth]{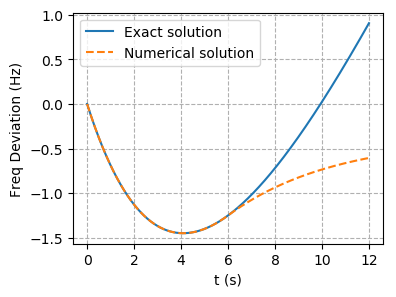} 
        \caption{6s ramp time} \label{fig:ramp_6s}
   \end{subfigure}
	\hfill
	\begin{subfigure}[t]{0.15\textwidth}
	\centering
        \includegraphics[width=\linewidth]{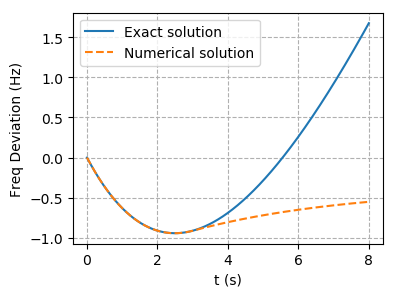} 
        \caption{3s ramp time} \label{fig:ramp_3s}
   \end{subfigure}
	\hfill
	\begin{subfigure}[t]{0.15\textwidth}
	\centering
        \includegraphics[width=\linewidth]{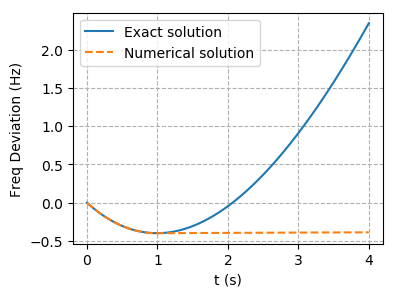} 
        \caption{1s ramp time} \label{fig:ramp_1s}
   \end{subfigure}
	\caption{Closed form vs numerical solutions - Linear ramp response}
	\label{fig:ramp_pfr}
\end{figure}

The deficiency of this formulation is that it is only accurate during the period of the ramp, i.e. $t \leq t_r$. The formulation does not allow the ramp to stabilise and flatten out (since the PFR equation $p(t)$ is non-differentiable if the maximum PFR quantity is applied as a hard limit after the ramp time is over). Figure \ref{fig:ramp_pfr} shows a comparison of the closed form solution against a numerically derived solution (with an integration step size of 1 ms)\footnote{The parameters used in this example are $P_{cont} = 300$ MW, $PFR = 270$ MW, $KE = 9,000$ MW.s, $P_{load} = 2,000$ MW and $D = 0.04$}. It can be seen that the frequency nadir is only accurately predicted when the nadir occurs before the ramp time is finished, i.e. Figures \ref{fig:ramp_6s} and \ref{fig:ramp_3s}.

In any case, this formulation can be extended to have an arbitrary number of response bands:

\begin{equation}
\Delta f(t) = \sum_{i \in N} \frac{R_i t}{D'} - \left[ \sum_{i \in N} \left( \frac{2 R_i H}{D'^{2}} \right) + \frac{P_{cont}}{D'} \right] \left( 1 - e^{-\frac{D'}{2H}t} \right)
\end{equation}

\noindent where $N$ is the number of response bands and $R_i = \frac{PFR_i}{t_{r,i}}$ is the ramp rate of the $i$-th band (in MW/s). However, it is clear that the accuracy of this extended formulation is limited to the response time of the fastest band. For applications that include FFR bands, then this formulation would not be very accurate at all (as can be seen in Figure \ref{fig:ramp_1s} where the closed form solution fails to predict the frequency nadir). This consideration is key when integrating new technologies with very fast ramps, e.g. inverter-interfaced systems such as batteries.

\subsection{Lag Response}
\label{sec:lag_response}
\begin{figure}
	\begin{center}
    \includegraphics[width=0.6\linewidth]{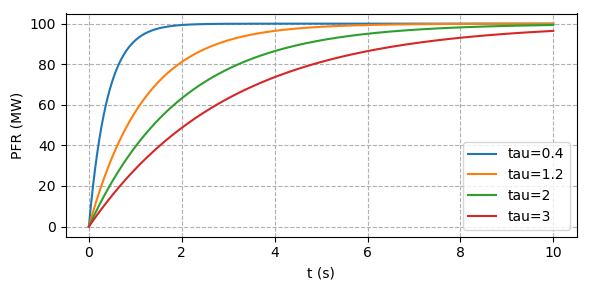}
    \end{center}
   	\caption{Comparison of lag PFR responses ($PFR = 100$ MW with different values of $\tau$)}
   	\label{fig:lag_pfr_compare}
\end{figure}

Consider a primary frequency response function of the form $p(t) = PFR ( 1 - e^{-t/\tau} )$, where $PFR$ is is the maximum quantity of PFR delivered (MW) and $\tau$ is a time constant governing the speed of response (s). Figure \ref{fig:lag_pfr_compare} graphically depicts how the parameter $\tau$ affects the speed of response.

The closed-form solution to (\ref{eqn:ode_form3}) for the lag response is:

\begin{equation}
\label{eqn:lag_pfr}
\begin{split}
    \Delta f(t) = \frac{PFR - P_{cont}}{D'} \left( 1 - e^{-\frac{D'}{2H}t} \right) \\ - \frac{PFR \times \tau}{D' \tau - 2H} \left( e^{-t/\tau} - e^{-\frac{D'}{2H}t} \right)
\end{split}
\end{equation}

\begin{figure}[!htp]
\begin{center}
\includegraphics[width=0.6\linewidth]{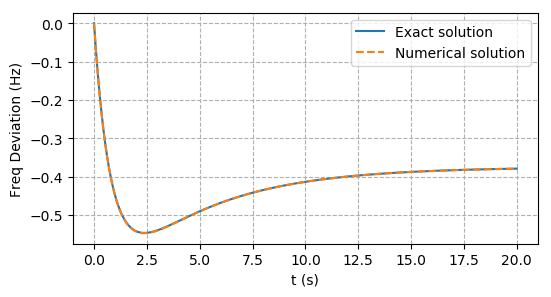}
\end{center}
\caption{Closed form vs numerical solution - Lag response}
\label{fig:lag_pfr}
\end{figure}

Unlike the linear ramp response, the lag response formulation is accurate for the whole simulation time. Figure \ref{fig:lag_pfr} shows a comparison of the closed form solution against a numerically derived solution (with an integration step size of 1 ms), indicating practically perfect alignment between the traces across the entire simulation time.

The time (in s) when the frequency nadir occurs can be solved by taking the derivative of (\ref{eqn:lag_pfr}) and setting it to zero:

\begin{equation}
\label{eqn:time_nadir}
t_{nadir} = \frac{\ln{ \left[ 1 + \frac{P_{cont}}{PFR} \left( \frac{D' \tau}{2H} - 1 \right) \right] }}{\frac{D'}{2H}-\frac{1}{\tau}}
\end{equation}

The frequency deviation (in Hz) at the nadir can then be calculated by inserting (\ref{eqn:time_nadir}) into (\ref{eqn:lag_pfr}). Note that (\ref{eqn:time_nadir}) can only be solved if: 

\begin{equation}
\label{eqn:PFR_constraint}
PFR >  P_{cont} \left( 1 - \frac{D' \tau}{2H} \right)
\end{equation}

When this condition does not hold, then the shape of the frequency response curve is asymptotic to the frequency nadir. This is because faster PFR responses tend to interact in concert with load relief effects resulting in a frequency that converges asymptotically to the nadir (see Figure \ref{fig:approx_vs_exact_a} as an example). In such cases, the frequency nadir can be solved by taking the limit of (\ref{eqn:lag_pfr}) as $t \to \infty$:
\begin{equation}
    \label{eqn:asymptotic_nadir}
    \begin{split}
    \Delta f_{nad} = \lim_{t\to\infty} \Delta f(t) = \frac{PFR - P_{cont}}{D'}
    \end{split}
\end{equation}

The maximum instantaneous rate of change of frequency (RoCoF) occurs at $t=0$:

\begin{equation}
    \label{eqn:max_rocof}
    \left( \frac{df}{dt} \right)_{max} = -\frac{P_{cont}}{2H}
\end{equation}

\subsection{Multiple Lag Response Bands}

The closed form solution in (\ref{eqn:lag_pfr}) can be readily extended to include an arbitrary number of response bands:

\begin{equation}
\label{eqn:multi_lag_pfr}
\begin{split}
    \Delta f(t) = \frac{\sum_{i \in N} PFR_i - P_{cont}}{D'} \left( 1 - e^{-\frac{D'}{2H}t} \right) \\ - \sum_{i \in N} \frac{ PFR_i \times \tau_i}{D' \tau_i - 2H} \left( e^{-t/\tau_i} - e^{-\frac{D'}{2H}t} \right)
\end{split}
\end{equation}

Unlike in (\ref{eqn:time_nadir}) for the lag response with only a single response band, there is no analytical solution for the roots of the derivative of (\ref{eqn:multi_lag_pfr}). Therefore, it would be desirable to develop an approximation for multiple response bands such that it would still fit in the single response band formulation.

\section{Approximation for Two Lag Response Bands}

Consider the primary frequency response function composed of two lag response bands:

\begin{equation}
p(t) = PFR_{1} ( 1 - e^{-t/\tau_1} ) + PFR_{2} ( 1 - e^{-t/\tau_2} )
\end{equation}

We want to find an approximate function for the sum of two exponentials in $p(t)$ that is formulated as a single exponential function, i.e. find equivalent parameters $PFR$ and $\tau$ such that:

\begin{equation}
\label{eqn:approx_exp}
PFR ( 1 - e^{-t/\tau} ) = PFR_{1} ( 1 - e^{-t/\tau_1} ) + PFR_{2} ( 1 - e^{-t/\tau_2} )
\end{equation}

\subsection{Canonical example}
\label{sec:canonical}

To further constrain the solution space and make this problem more tractable, let us consider a practical example and assume for instance the speed of response parameters $\tau_{1} = 0.4$ and $\tau_{2} = 2.0$ (see Figure \ref{fig:lag_pfr_compare}), corresponding to a fast (90\% of full response in 1s) and moderate-speed (90\% of full response in 5s) response respectively.

\begin{figure}[!thp]
	\centering
	\begin{subfigure}[t]{0.225\textwidth}
	\centering
        \includegraphics[width=\linewidth]{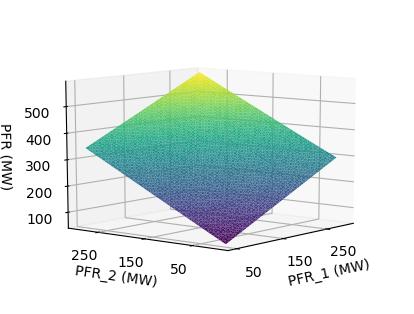} 
        \caption{$PFR$} \label{fig:approx_exp_a}
   \end{subfigure}
	\hfill
	\begin{subfigure}[t]{0.225\textwidth}
	\centering
        \includegraphics[width=\linewidth]{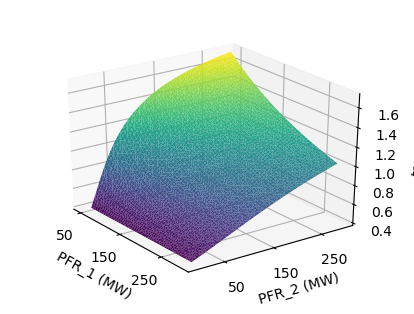} 
        \caption{$\tau$} \label{fig:approx_exp_b}
   \end{subfigure}
\caption{Curve fitting results for parameters $PFR$ and $\tau$}
\label{fig:approx_exp}
\end{figure}

A non-linear least squares algorithm (trust region reflective method \cite{byrd_1988}) was used to fit the sum of exponentials with the equivalent approximation in (\ref{eqn:approx_exp}) and find fitted values of $PFR$ and $\tau$ for a range of values of $PFR_1$ and $PFR_2$. 

The results of the curve fitting exercise are shown in the pair of 3D surface plots in Figure \ref{fig:approx_exp}. It can be seen from Figure \ref{fig:approx_exp_a} that the parameter $PFR$ can be approximated as a linear plane and thus $PFR \approx PFR_1 + PFR_2$. However, the parameter $\tau$ is clearly a non-linear surface and needs further approximation. A Levenberg-Marquardt (LM) curve fitting algorithm \cite{marquardt_1963} was used to fit the data in Figure \ref{fig:approx_exp_b} to a model equation of the form:

\begin{equation}
\label{eqn:approx_tau}
\tau = a \left[ 1 - e^{-b \left( \frac{PFR_2}{PFR_1} \right)} \right] + \tau_{1}
\end{equation}

\begin{figure}[!htp]
\begin{center}
\includegraphics[width=0.6\linewidth]{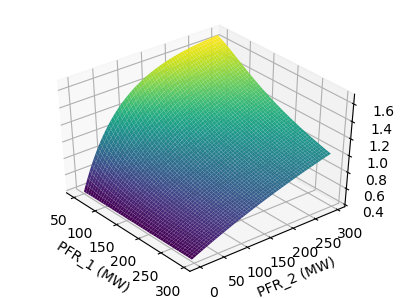}
\end{center}
\caption{Approximate equation for $\tau$ as per $PFR_1$ and $PFR_2$}
\label{fig:approx_tau}
\end{figure}

The LM algorithm resulted in the coefficients $a = 1.3141629$ and $b = 0.63075533$. Using these coefficients in the model equation (\ref{eqn:approx_tau}) yields the surface plot in Figure \ref{fig:approx_tau}, which is fairly representative of the shape and values in Figure \ref{fig:approx_exp_b}. 

To summarise, in this canonical example we can represent a PFR with two lag response bands with speed of response parameters $\tau_1=0.4$ (fast) and $\tau_1=2.0$ (standard) as an equivalent single lag response:

\begin{equation}
\hat{p}(t) = PFR' ( 1 - e^{-t/\tau'} )
\end{equation}
\begin{equation}
PFR' = PFR_1 + PFR_2
\end{equation}
\begin{equation}
\tau' = 1.3141629 \left[ 1 - e^{-0.63075533 \left( \frac{PFR_2}{PFR_1} \right)} \right] + 0.4
\end{equation}

\noindent where $PFR_1$ is the fast response (MW) and $PFR_2$ is the standard response (MW).

\subsection{Accuracy of the approximation}

\begin{figure}[!htp]
	\centering
	\begin{subfigure}[t]{0.225\textwidth}
	\centering
        \includegraphics[width=0.95\linewidth]{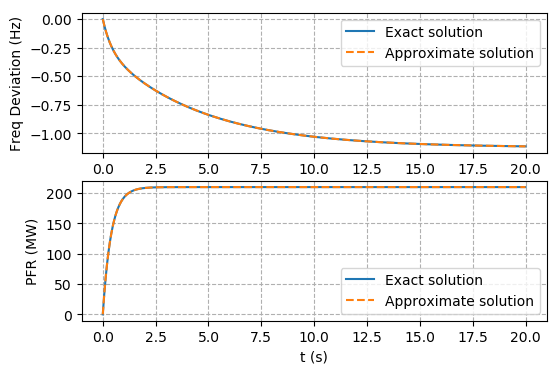} 
        \caption{$PFR_{1} = 210$ MW, $PFR_{2} = 0$ MW}
		  \label{fig:approx_vs_exact_a}
   \end{subfigure}
	\hfill
	\begin{subfigure}[t]{0.225\textwidth}
	\centering
        \includegraphics[width=0.95\linewidth]{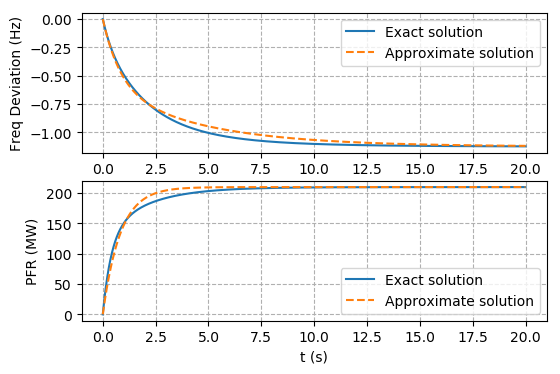} 
        \caption{$PFR_{1} = 130$ MW, $PFR_{2} = 80$ MW} 
		  \label{fig:approx_vs_exact_b}
   \end{subfigure}
	\vfill
	\begin{subfigure}[t]{0.225\textwidth}
	\centering
        \includegraphics[width=0.95\linewidth]{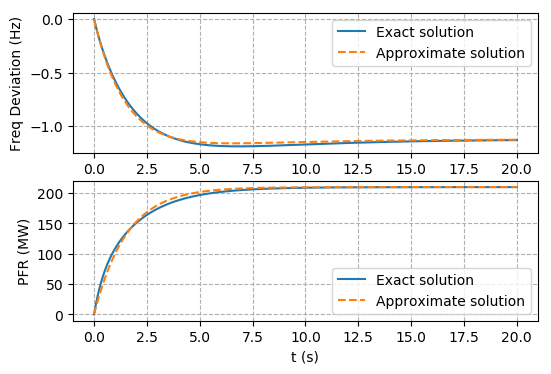} 
        \caption{$PFR_{1} = 50$ MW, $PFR_{2} = 160$ MW} 
		  \label{fig:approx_vs_exact_c}
   \end{subfigure}
	\hfill	
	\begin{subfigure}[t]{0.225\textwidth}
	\centering
        \includegraphics[width=0.95\linewidth]{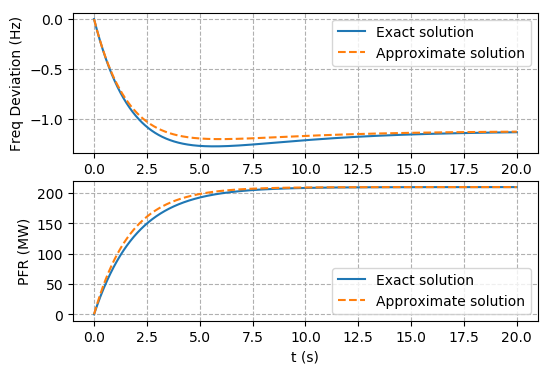} 
        \caption{$PFR_{1} = 0$ MW, $PFR_{2} = 210$ MW} 
		  \label{fig:approx_vs_exact_d}
   \end{subfigure}
	\caption{Exact vs approximate closed form solutions}
	\label{fig:approx_vs_exact}
\end{figure}

Figure \ref{fig:approx_vs_exact} shows a series of comparisons between the exact closed form solution with two lag response bands and the approximate solution with a single equivalent response band, for varying quantities of $PFR_1$ and $PFR_2$. The system conditions for these plots are: $P_{cont} = 300$ MW, $KE = 9,000$ MW.s, $P_{load} = 2,000$ MW and $D = 0.04$. 

It can be seen from the plots that the approximation is exact when only $PFR_1$ is used, and has the highest errors when only $PFR_2$ is used.

The accuracy of the approximation can be measured by calculating the Mean Absolute Percentage Error (MAPE):

\begin{equation}
    \label{eqn:MAPE}
    MAPE = \frac{1}{N} \sum_{i=1}^{N} \left| \frac{p(i) - \hat{p}(i)}{p(i)} \right| \times 100
\end{equation}

\noindent where $p$ is the exact solution, $\hat{p}$ is the approximate solution and $N$ is the number of samples.

\begin{figure}[!htp]
\begin{center}
\includegraphics[width=0.8\linewidth]{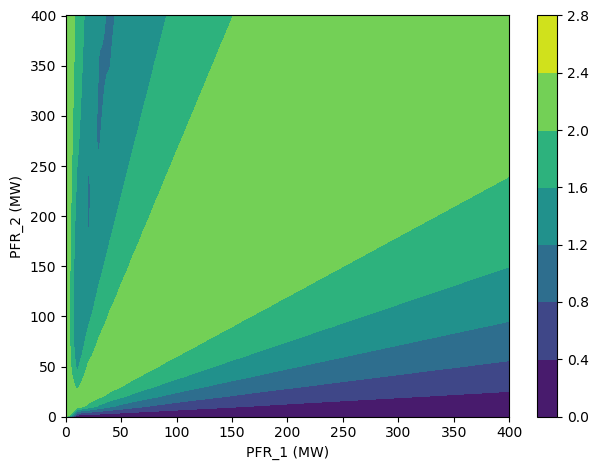}
\end{center}
\caption{MAPE values for the canonical example with different values of $PFR_1$ and $PFR_2$}
\label{fig:MAPE_contour}
\end{figure}

The MAPE values for different values of $PFR_1$ and $PFR_2$ in the canonical example are plotted on a contour map in Figure \ref{fig:MAPE_contour}. The MAPE values confirm the earlier observations from Figure \ref{fig:approx_vs_exact}, whereby the errors are minimised at higher proportions of $PFR_1$ relative to $PFR_2$. The mean MAPE in the contour map is 1.8\%, while the maximum MAPE is 2.4\%.

\subsection{Validity of the approximation for other $\tau_1$ and $\tau_2$ values}

\begin{figure}[!htp]
\begin{center}
\includegraphics[width=1.0\linewidth]{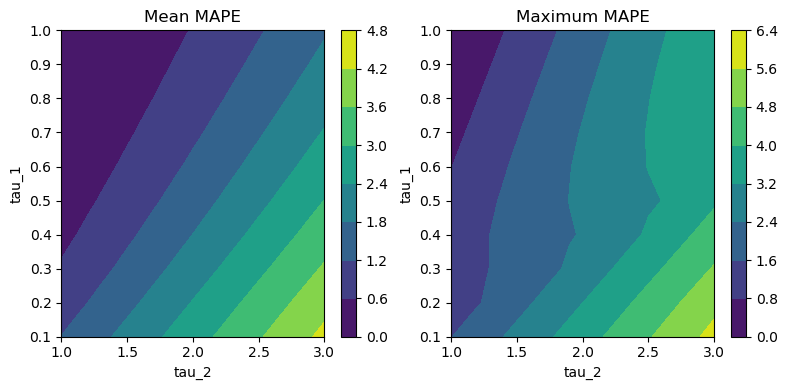}
\end{center}
\caption{Mean and maximum MAPE values for different values of $\tau_{1}$ and $\tau_{2}$}
\label{fig:mean_max_MAPE}
\end{figure}

Figure \ref{fig:mean_max_MAPE} shows contour maps of the mean and maximum MAPE values for a range of speed of response parameters $\tau_1$ and $\tau_2$ that are different from the canonical example. It can be seen that the errors increase when the ratio of $\tau_2$ to $\tau_1$ is large, and converge to zero as the ratio draws closer to unity. The mean MAPE and maximum MAPE across the range of $\tau_1$ and $\tau_2$ values are 1.58\% and 5.8\% respectively. The generally low error values ($<$5\%) suggest that the proposed approximation method is broadly valid across a range of $\tau_{1}$ and $\tau_{2}$ values.

\section{Example Applications}
\subsection{Example 1: Maximum allowable contingency size}
The frequency deviation (in Hz) at the nadir for the lag response, calculated by inserting (\ref{eqn:time_nadir}) into (\ref{eqn:lag_pfr}), can be formulated as follows (refer to Appendix \ref{sec:appendix2} for the proof):

\begin{equation}
    \label{eqn:lag_nadir}
    \Delta f_{nad} = \frac{PFR}{D'} \left[ (C + K - 1) B^{-C} - C B^{-C/A} - K + 1 \right]
\end{equation}

\noindent where $\frac{1}{K} = \frac{PFR}{P_{cont}}$ represents the ratio of the total PFR to the contingency size, $A = \frac{D' \tau}{2H}$, $B = 1 + K (A - 1)$ and $C = \frac{A}{A-1}$.

In jurisdictions such as the WEM, a minimum PFR constraint is imposed that is relative to the size of the largest contingency, which can be expressed as follows:

\begin{equation}
    \label{eqn:K_factor}
    PFR \geq \frac{P_{cont}}{K}
\end{equation}

For example, the minimum PFR in the WEM must be at least 70\% of the largest contingency \cite{WEM_2020}. Therefore in this case, $K$ is a constant, i.e. $K = \frac{1}{0.7} = 1.429$.

Given a maximum allowable frequency deviation $\Delta f_{max}$, the maximum contingency size allowed (while exactly meeting the PFR constraint) can be calculated as follows:

\begin{equation}
\label{eqn:max_contingency}
    P_{cont} \leq \frac{K \times D' \Delta f_{max}}{(C + K - 1) B^{-C} - C B^{-C/A} - K + 1}
\end{equation}

Using this formulation on the canonical example of the two lag response band approximation (described in Section \ref{sec:canonical}), the maximum allowable contingency size can be calculated directly for different values of $\tau$, and then by extension, the corresponding minimum proportion of FFR required to maintain system security. 

\begin{figure}[!htp]
\begin{center}
\includegraphics[width=1.0\linewidth]{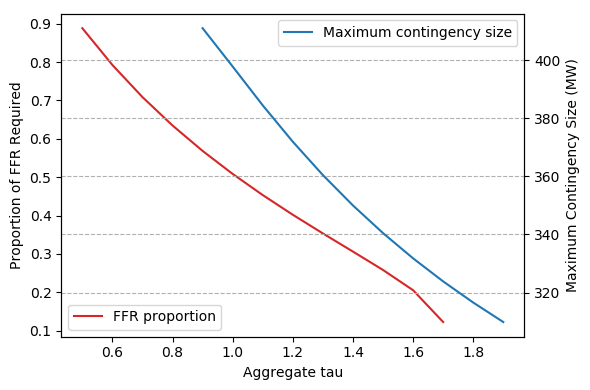}
\end{center}
\caption{Maximum allowable contingency size for different values of $\tau$}
\label{fig:max_contingency}
\end{figure}

An example is shown in Figure \ref{fig:max_contingency}, where the conditions are: $\Delta f_{max} = -1.25$ Hz, $KE = 7,000$ MW.s, $P_{load} = 2,500$ MW and $D = 0.04$. From this plot, it can be seen that high contingency sizes require lower values of $\tau$, and correspondingly high proportions of FFR. For instance, a maximum contingency size of 400 MW needs an aggregate $\tau$ value of roughly 1.0, which in turn requires roughly 51\% of FFR in the total PFR mix. 

An observation that can be made from (\ref{eqn:max_contingency}) is that the maximum allowable contingency size can be expressed as an equation of $A$, $K$ and $D'$:

\begin{equation}
    P_{cont} \leq f(A, K) D'
\end{equation}
\noindent where 
\begin{equation*}
    f(A, K) = \frac{K \times \Delta f_{max}}{(C + K - 1) B^{-C} - C B^{-C/A} - K + 1}
\end{equation*}

Note that $f(A, K)$ can only be solved if $A \geq 1 - \frac{1}{K}$. If this inequality does not hold, then as noted earlier in Section \ref{sec:lag_response}, the shape of the frequency response curve is asymptotic to the frequency nadir. In such cases, the maximum allowable contingency can be determined by re-arranging (\ref{eqn:asymptotic_nadir}):

\begin{equation}
    P_{cont} \leq \frac{\Delta f_{max}}{\frac{1}{K} - 1} D'
\end{equation}

If the maximum allowable contingency was expressed as a multiple of $D'$ (which is a function of system load and damping factor), then universally applicable metrics can be obtained in terms of $A$ and $K$ (since $A = \frac{D' \tau}{2H}$ and $K = \frac{P_{cont}}{PFR}$ are expressed as ratios and not absolute values).

\begin{figure}[!htp]
\begin{center}
\includegraphics[width=1.0\linewidth]{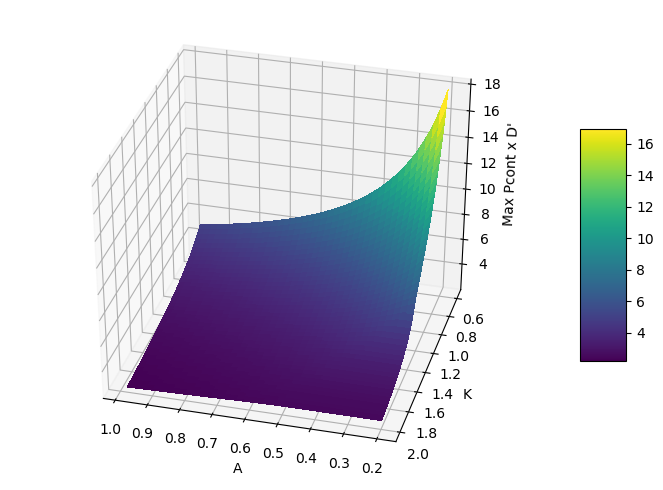}
\end{center}
\caption{Universally applicable maximum allowable contingency in terms of the metrics $A$ and $K$}
\label{fig:max_contingency_Dp}
\end{figure}

This is shown as a 3D plot in Figure \ref{fig:max_contingency_Dp}, which provides fundamental insight into how the maximum allowable contingency changes with respect to changes in the relative ratios of $\tau$ to $H$ and $P_{cont}$ to $PFR$. As expected, the maximum allowable contingency size increases when either $A$ is reduced (e.g. by increasing $H$ or decreasing $\tau$) or $K$ is reduced (e.g. by increasing the amount of $PFR$ relative to the contingency size).

\subsection{Example 2: Practical lower bound to $\tau$}

The inequality in (\ref{eqn:PFR_constraint}) can be rearranged to be in terms of $\tau$:
\begin{equation}
    \label{eqn:min_tau}
    \tau \geq \left( 1 - \frac{1}{K} \right) \frac{2H}{D'}
\end{equation}

It can be seen from (\ref{eqn:asymptotic_nadir}) that when this inequality doesn't hold, the frequency nadir is asymptotic and independent of $\tau$. The implication of this is that reducing $\tau$ below the threshold in (\ref{eqn:min_tau}) has no effect on the resulting frequency nadir. Therefore, (\ref{eqn:min_tau}) sets a practical lower bound for $\tau$ at which point no further performance benefits are seen (vis-a-vis the frequency nadir) by increasing the aggregate speed of PFR response.

\subsection{Example 3: Relative trade-offs between contingency size, PFR,  system inertia and aggregate speed of response}

Consider the special case where the PFR is equal to the contingency size, i.e. $K = 1$. In this special case, the expression for the maximum contingency size in (\ref{eqn:max_contingency}) reduces to:

\begin{equation}
\label{eqn:max_contingency_special}
    P_{cont} = PFR \leq - \frac{D' \Delta f_{max}}{A ^ {-\frac{1}{A-1}}}
\end{equation}

Since $A$ is defined as a function of $H$ and $\tau$, the sensitivity of the contingency size $P_{cont}$ to changes in system inertia and aggregate speed of response can be calculated by finding the following partial derivatives:

\begin{equation}
    \frac{\partial P_{cont}}{\partial \tau} = - \frac{D' \Delta f_{max}}{\tau} \left[ \frac{A - 1 - A \ln{A}}{ \left( A - 1 \right)^{2}} \right] A ^ {-\frac{1}{A-1}}
\end{equation}

\begin{equation}
    \frac{\partial P_{cont}}{\partial H} = \frac{D' \Delta f_{max}}{H} \left[ \frac{A - 1 - A \ln{A}}{ \left( A - 1 \right)^{2}} \right] A ^ {-\frac{1}{A-1}}
\end{equation}

Given the approximation for $\tau$ in (\ref{eqn:approx_tau}) for a PFR with two response bands, the sensitivity of $\tau$ to changes in $PFR_1$ and $PFR_2$ are:

\begin{equation}
    \frac{\partial \tau}{\partial PFR_{1}} = -ab \frac{PFR_{2}}{(PFR_{1})^{2}} e^{-b \left( \frac{PFR_2}{PFR_1} \right)}
\end{equation}

\begin{equation}
    \frac{\partial \tau}{\partial PFR_{2}} = \frac{ab}{PFR_1} e^{-b \left( \frac{PFR_2}{PFR_1} \right)}
\end{equation}

$\frac{\partial P_{cont}}{\partial PFR_1}$ and $\frac{\partial P_{cont}}{\partial PFR_2}$ can then be readily calculated using the chain rule.

If marginal prices were known for the cost of changing the contingency size by 1 MW, the aggregate speed of response $\tau$ (or, by extension, the relative proportions of $PFR_1$ and $PFR_2$) and if possible, the system inertia by 1 MW.s, then the sensitivities could be applied to arrive at an optimal least cost mix of parameters. Note that in its present form, such a solution would not be co-optimised with energy dispatch or other ancillary services.

The approach described above for analysing sensitivities can also be generalised for all values of $K$ by taking partial derivatives of (\ref{eqn:max_contingency}) with respect to $K$, i.e. $\frac{\partial P_{cont}}{\partial K}$ and then using the chain rule to find $\frac{\partial P_{cont}}{\partial PFR}$.

\section{Conclusion}
This paper presented a novel closed-form solution for a low-order SFR model that is accurate for all time periods and an approximation of the PFR that can support two different speeds (e.g. a fast response and standard response) while still maintaining mathematical tractability, i.e. there exist closed-form expressions for the frequency nadir and RoCoF. 

The proposed closed-form solutions are useful for practical applications that will benefit from direct computation of the frequency nadir or RoCoF (in lieu of numerical solutions), while still taking into account the influence of fast frequency responses. For example, existing security-constrained unit commitment solutions considering frequency stability limits (such as in \cite{Ahmadi_2014}) can potentially be extended to include fast frequency responses.

An avenue for future work is the consideration of PFR activation delays in the formulation. The inclusion of activation delays may be significant for very fast PFR responses (e.g. batteries) where the activation delay may end up being of greater duration than the PFR response time itself. 

\appendices

\section{Proof of the lag response frequency nadir}
\label{sec:appendix2}

The time when the frequency nadir occurs in (\ref{eqn:time_nadir}) can be expressed as follows:
\begin{equation}
    \label{eqn:time_nadir_simplified}
    t_{nadir} = \frac{\tau \ln{B}}{A - 1}
\end{equation}
\noindent where $K = \frac{P_{cont}}{PFR}$, $A = \frac{D' \tau}{2H}$ and $B = 1 + K (A - 1)$.

Inserting (\ref{eqn:time_nadir_simplified}) into (\ref{eqn:lag_pfr}) and simplifying:

\begin{equation}
    \Delta f_{nad} = \frac{PFR}{D'} \left[ (C + K - 1) B^{-C} - C B^{-C/A} - K + 1 \right]
\end{equation}
\noindent where $C = \frac{D' \tau}{D' \tau - 2H} = \frac{A}{A-1}$.


\bibliographystyle{IEEEtran}
\bibliography{bib_refs}

\end{document}